\documentstyle[preprint,aps,epsfig]{revtex}   
\def\beq{\begin{equation}}
\def\eeq{\end{equation}}
\def\be{\begin{eqnarray}}
\def\ee{\end{eqnarray}}
\def\ci{\cite}
\def\bi{\bibitem}

\def\magp{|{\bf p}|}            
\begin{document}

\draft

\title{ Nuclear structure functions and off-shell corrections }

\author
{ O. Benhar$^{1}$,  S. Fantoni${^2}$, G.I. Lykasov${^3}$
\cite{byline} }

\address
{ $^1$ INFN, Sezione Roma 1, I-00185 Rome, Italy \\
$^2$ International School for Advanced Studies (SISSA)\\
and \\INFN, Sezione di Trieste. I-30014 Trieste, Italy\\
$^3$ Joint Institute for Nuclear Research, Dubna 141980, Moscow
Region, Russia }

\date{\today}

\maketitle

\begin{abstract}

The $n$-th moments of the nuclear structure function $F_2^A(x,Q^2)$
are analyzed using the off-shell kinematics appropriate to 
describe deep inelastic electron-nucleus scattering within the 
impulse approximation. It is shown that off-shell effects are 
sensitive to the form of both the nuclear spectral function and 
the nucleon structure function $F_2^N(x,Q^2)$, and can be as large
as $\sim 10\%$ at Q$^2~\sim~2$~(GeV/c)$^2$.
\end{abstract}
  

Nuclear effects in deep inelastic
lepton-nucleus scattering (DIS) have been extensively studied, 
both experimentally and theoretically, over the last two decades 
(for a review see refs. \ci{arneodo,geesaman}).
The body of available data provides clearcut evidence that 
the nucleus cannot be simply described as a collection 
of nucleons on the mass shell. 
Proton knock-out experiments in which the outgoing particle is detected
in coincidence with the scattered electron 
have clearly shown that the nuclear spectral function 
$P({\bf p},E)$, yielding the probability
of finding a nucleon with momentum ${\bf p}$ and removal energy $E$ 
in the target nucleus, extends to very large values of $\magp$ 
($\gtrsim$ 800 MeV/c) and $E$ ($\gtrsim$ 200 MeV) \ci{vj}. 
Hence, the theoretical description of 
lepton scattering off fast off-shell nucleons must be regarded as a 
prerequisite for the understanding of nuclear DIS data. 

A pioneering study of the electromagnetic response of a bound nucleon, 
focused on the intermediate energy domain, has been carried out in the 80's 
by de Forest \ci{forest}.
More recent theoretical analyses of off-shell effects in nuclear DIS 
\ci{sim1,rin1} have been stimulated by the availability of 
new experimental information on the $Q^2$-dependence of the moments of the 
nuclear structure function $F_2^A(x,Q^2)$, showing sizable deviations from 
the prediction of the simple convolution model \ci{sim1}.
   
In this paper we study the relevance of nuclear effects to the $Q^2$-dependence
of the nuclear structure function $F_2^A(x,Q^2)$ and its moments using 
the appropriate off-shell kinematics and a realistic nuclear spectral function,
resulting from {\it ab initio} microscopic many-body calculations \ci{bff}.
    

Let us consider electron scattering off a nucleon carrying
four momentum $p \equiv (p_0,{\bf p})$ with $p^2 \ne m^2$, $m$ being 
the nucleon mass.  As it is well known (see, e.g., ref.\cite{close}), the 
differential 
cross section for the inclusive process $e~+~N~\rightarrow~e^\prime~+~X$, 
in which the hadronic final state is undetected, can be written in 
the form 
\beq
\frac{d^2\sigma_N}{d\Omega dE_{e^\prime}} = \frac{1}{{\rm F}(k,p)}\ 
\frac{\alpha^2}{Q^4} \frac{E_{e^\prime}}{E_e} L^{\mu\nu}W_{\mu\nu}\ ,
\label{dis:dcs} 
\eeq
where $\alpha^2=e^2/4\pi=1/137$, $q^2 = \nu^2 - |{\bf q}|^2 = -Q^2$ is the 
squared four-momentum tranfer and $E_e$ and $E_{e^\prime}$ are the initial 
and final electron energy, respectively. The function 
${\rm F}(k,p)~=~v_{rel}~=~|({\bf k}/E_e)-({\bf p}/p_0)|$, $k \equiv (E_e,{\bf k})$ 
being the four momentum
carried by the incoming electron, describes the incident flux. In the case of
ultrarelativistic electrons it takes the simple form 
${\rm F}(k,p) = (kp)/(p_0E_e)$. 
Note that for an isolated stationary nucleon $F \equiv 1$.

The electromagnetic structure of the electron and the
nucleon is described by the tensors $L^{\mu\nu}$ and $W_{\mu\nu}$, respectively. 
Expanding the hadronic tensor in terms of the available relativistic 
invariants and taking into account current conservation, the contraction 
$L^{\mu\nu}W_{\mu\nu}$ can be readily written in terms of two nucleon
structure functions $W_1$ and $W_2$:
\beq
L^{\mu\nu}W_{\mu\nu} = 2 \left\{ 2\ W_1 (k k^\prime)
+ \frac{W_2}{m^2} \left[ 2(k p)(k^\prime p) - p^2 (k k^\prime)
\right] \right\}\ ,
\label{contr:1}
\eeq
where $k^\prime = k + q$ denotes the four momentum carried by the 
 outgoing electron and 
$(k k^\prime) = 2 E_{e}E_{e^\prime} \sin^2(\theta/2)$, $\theta$ being the 
electron scattering angle. In the case of scattering off a free nucleon at rest 
$p^2 = m^2$, $(k p)(k^\prime p) = m^2 E_{e} E_{e^\prime}$ and
substitution of eq.(\ref{contr:1}) into eq.(\ref{dis:dcs}) yields that 
standard result
\beq
\frac{d^2\sigma_N}{d\Omega dE_{e^\prime}} = 
\sigma_{M} \left[ W_2 + 2\ W_1 \tan^2 \frac{\theta}{2} \right]\ ,
\label{def:dcs1}
\eeq
where $\sigma_{M}=\alpha^2 \cos^2(\theta/2)/[4 E_{e}^2 \sin^4(\theta/2)]$
is the Mott cross section and the nucleon structure functions depend upon
$q^2$ and $(qp)$ only. In general, the inclusive electron-nucleon cross-section
contains additional terms and can be cast in the form
\beq
\frac{d^2\sigma_N}{d\Omega dE_{e^\prime}} =
\frac{\sigma_{M}}{{\rm F}(k,p)} \left[ W_2 \left( 1 - \Delta \right) 
+ 2\ W_1 \tan^2 \frac{\theta}{2} \right]\ ,
\label{def:dcs2}
\eeq
with
\beq
\Delta(k,k^\prime,p) = \left[ 1 - 
\frac{(k p)}{m E_{e}}\frac{(k^\prime p)}{m E_{e^\prime}} 
\right]
 \cos^{-2} \frac{\theta}{2} - \left( 1 - \frac{p^2}{m^2} \right) \tan^2 
\frac{\theta}{2}\ ,
\label{def:delta}
\eeq
and the structure functions $W_1$ and $W_2$ depend upon the three relativistic
invariants $q^2$, $(qp)$ and $p^2$. For an isolated stationary nucleon $p^2 = m^2$, 
${\rm F}(k,p)=1$, $\Delta(k,k^\prime,p) = 0$ and eq.(\ref{def:dcs1}) is recovered.

\clearpage

The electron-nucleon cross section of eq.(\ref{def:dcs2}) can be used to obtain 
the inclusive electron-nucleus cross section within the impulse approximation 
from \ci{bfls2}:
\beq
\frac{d\sigma_A} {d\Omega d E_{e^\prime}} =
\int d^4 p \left (\frac{m}{p_0} \right) P(p)\ \left[ {\rm F}(k,p)\  
\left( \frac{d^2\sigma_N}{d\Omega dE_{e^\prime}} \right) \right]
\label{def:dcsa}
\eeq
where $P(p) \equiv P(\magp,m-p_0)$ denotes the nuclear spectral function. 
The interpretation of the above equation is straightforward. The 
electron-nucleus cross section is written as the incoherent sum of the
cross sections associated with scattering off individual nucleons, 
weighted with the nuclear spectral function and stripped of the flux factors, 
since the incident flux has to be defined with respect to the
target, i.e. to the  nucleus as a whole.
Note that the nuclear cross section of eq.(\ref{def:dcsa}) can be recast in a form 
similar to eq.(\ref{def:dcs1}), with the nucleon structure functions $W_1$ and $W_2$
replaced by two nuclear structure functions $W_1^A$ and $W_2^A$.

Due to the fact that $P(p)$ is a steeply decreasing function of 
$|{\bf p}|$, the main contribution to the $|{\bf p}|$ integration involved in 
eq.(\ref{def:dcsa}) comes from the region of ${\bf p}^2$ small compared 
to $m^2$, where the use of a nonrelativistic nuclear spectral
function is quite reasonable. Furthermore, as a first approximation, 
 the functions $W_1$ and $W_2$, whose $p^2$-dependence is unknown, can be estimated 
assuming $W_{1,2}(q^2,(qp),p^2)\simeq W_{1,2}(q^2,(qp),m^2)$, and 
$W_{1,2}(q^2,(qp),m^2)$ can be extracted from electron-proton and electron-deuteron 
scattering data.

Using eqs.(\ref{def:dcs1}) and (\ref{def:dcsa}) the nuclear structure function
$F_2^A=\nu W_2^A$ can be related to the nucleon structure function $F_2=\nu W_2$ 
through
\beq
F_2^A(x,Q^2)=\int_x^A dz\ \phi_0(z,Q^2) F_2 \left( \frac{x}{z},Q^2 \right)\ ,
\label{def:f2aof}
\eeq
where
\beq
\phi_0(z,Q^2)  =  z \int d^4p\ \left (\frac{m}{p_0} \right) P(p)\  
 \left[ 1 - \Delta(k,k^\prime,p) \right] \ 
\delta \left(z-\frac{(p q)}{(p_A q)}\frac{M_A}{m} \right)\ ,
\label{def:phi0}
\eeq
$M_A$ being the mass of the target nucleus.

Note that setting $\Delta$ = 0  eq.(\ref{def:f2aof}) reduces to the standard
convolution formula (see, e.g., ref \ci{bfls1}):  
\beq
{\widetilde F}_2^{A}(x,Q^2) =  \int_x^A dz f_A(z,Q^2) 
F_2 \left(\frac{x}{z},Q^2 \right) \ ,
\label{def:f2a}
\eeq  
with
\beq
f_A(z,Q^2)= z \int d^4p \left( \frac{m}{p_0} \right) P(p)\  
\delta \left(z - \frac{(pq)}{(P_Aq)}\frac{M_A}{m} \right)\ .
\label{def:fax1}
\eeq

Let us now focus on the calculation of the $n$-th moment of the nuclear structure 
function, defined as
\beq
M_n^A(Q^2) = \int_0^A dx\  x^{n-2} F_2^A(x,Q^2)\ .
\label{def:moma}
\eeq
Using the convolution model result of eq.(\ref{def:f2a}) and assuming the
validity of the Bjorken limit, which amounts to disregard the 
$Q^2$-dependence of $f_A(z,Q^2)$ of eq.(\ref{def:fax1}), eq.(\ref{def:moma})
can be rewritten in the simplified form
\beq
{\widetilde M}_n^{A}(Q^2) = \int_0^A dz\int_0^1 dy\ z^{n-1} f_A(z)
y^{n-2} F_2 \left(y,Q^2 \right)\ ,
\label{def:momon}
\eeq
leading to the factorized expression
\beq
{\widetilde M}_n^{A}(Q^2) = f_A^{(n+1)} M_n^N(Q^2)\ .
\label{def:fact}
\eeq
In the above equation $M_n^N(Q^2)$ is the n-th nucleon moment and
\beq
f_A^{(n+1)} = \int_0^A dz\ z^{n-1}\ f_A(z)\ ,
\eeq 
whereas $f_A(z)$ denotes the $Q^2 \rightarrow \infty$ limit of 
$f_A(z,Q^2)$ \ci{bfls1}.

Substitution of the nuclear structure function given by 
eqs.(\ref{def:f2aof}) and (\ref{def:phi0}) into eq.(\ref{def:moma})
 shows that the factorization property exhibited by 
the n-th moment evaluated within the convolution model breaks
down when the Fermi motion and binding of the struck nucleon are
properly taken into acount. 

The breaking of factorization can be best observed studying the ratio 
between the nuclear and nucleon n-th moments, defined as  
\beq
R_n(Q^2) = \frac {M_n^A(Q^2)}{M_n^N(Q^2)} \ .
\label{def:Rn}
\eeq
According to the convolution model, $R_n(Q^2)$ is $Q^2$-independent, hence
the study of its $Q^2$ dependence may provide useful information on the
relevance of the factorization breaking terms arising from the inclusion 
of off-shell kinematics for the struck nucleon in the calculation 
of $F_2^A$ from eqs.(\ref{def:f2aof})-(\ref{def:phi0}).

The results of our numerical investigation of the effects of factorization 
breaking on the ratio $R_5(Q^2)$ are shown in fig.\ref{fig:1}.
The solid curve has been obtained using $F_2^A(x,Q^2)$ given by 
eqs.(\ref{def:f2aof}) and (\ref{def:phi0}) and the nuclear matter 
spectral function of ref.\cite{bff}, whereas the data points, taken 
 from ref.\ci{sim1}, have been extracted from the analysis of 
several experiments carried out at CERN \ci{cern} and SLAC \ci{slac1,slac2}
using $^{56}$Fe targets. 
It appears that the calculated $R_5(Q^2)$ decreases in the range 
$2 \lesssim Q^2 \lesssim 10$ (GeV/c)$^2$, while becoming almost constant at 
$Q^2 >$ 10 (GeV/c)$^2$. The maximum deviation from the prediction of
the convolution model, shown by the horizontal line in fig. \ref{fig:1},
is $\sim 15 \%$. Our results appreciably depend upon the form of the 
nucleon structure function $F_2(x,Q^2)$, particularly at 
$Q^2 <$ 10 (GeV/c)$^2$. In this region we have employed the 
parametrization proposed in refs. \ci{ck1,ck2}, generally referred to 
as CKMT model, that provides a consistent formulation
of the structure function at low $Q^2$ and gives a good description of 
the recent data from HERA \ci{hera1,hera20,hera21,hera3,hera4}.
According to the CKMT model, based on Regge theory, the nucleon structure 
function $F_2^N$ can be split into two parts: the singlet term, corresponding 
to the Pomeron contribution, and the nonsinglet term, corresponding to the 
secondary Reggeon. Both terms have the nonpertubative $Q^2$-dependence. 
At $Q^2\geq$~10~(GeV/c)$^2$ we have used the so called GRV model, proposed in  
ref. \ci{grv}, which reproduces the $Q^2$-evolution of $F_2(x,Q^2)$
quite well.
 
We have also used our approach to calculate the quantity
\beq
\tau^A_n(Q^2) = 
\left[ \frac{M_n^A(Q^2)}{M_n^A(Q_0^2)} \right]^{-(1/d_n)}\ , 
\label{def:tau}
\eeq
where $d_n=\gamma_n/2\beta_0$, $\gamma_n$ is the so called non singlet (NS) 
anomalous dimension and $\beta_0~=$~11$-$(2/3)$n_f$, $n_f$ being the number 
of flavors. 

The solid line of fig. \ref{fig:2} shows $\tau^A_5(Q^2)$, obtained 
from eqs.(\ref{def:f2aof})-(\ref{def:phi0}) and (\ref{def:moma}), for 
$Q_0^2 =$ 12.5 (GeV/c)$^2$.  It exhibits a deviation 
from the $\log(Q^2/\Lambda^2)$ behavior predicted 
by leading order perturbative QCD and the factorization theorem.
This deviation is a consequence of {\em both} higher twist 
corrrections, included in the CKMT parametrization of the nucleon structure
function, {\em and} nuclear effects. In order to illustrate the relative 
importance of the two sources of non logarithmic $Q^2$-dependence of 
$\tau^A_5(Q^2)$, we also show, by the dashed line, the results obtained 
setting $\Delta =$ 0 in eq.(\ref{def:phi0}). Comparison between the solid 
and dashed lines shows that nuclear effects reach a maximum of $\sim$ 10 \%
at $Q^2=2 (GeV/c)^2$.

The results presented in figs. \ref{fig:1} and \ref{fig:2} can be
related to each other noting that, defining the nucleon analog 
of $\tau^A_5(Q^2)$ as $\tau^N_5(Q^2)=[M_5^N(Q^2)/M_5^N(Q_0^2)]^{-1/d_5}$, 
we can write
\beq
\frac{\tau_5^N(Q^2)}{\tau_5^A(Q^2)} = 
\left[\frac{ R_5(Q^2)}{R_5(Q_0^2)} \right]^{1/d_5}\ .
\label{def:link}
\eeq
Numerical calculations show that $\tau^N_5(Q^2)$ is very close 
to the quantity represented by the dashed line of fig. \ref {fig:2}. 
 Hence, to a very good approximation, the above equation provides a 
relationship between 
$R_5(Q^2)$ of fig. \ref{fig:1} and the ratio of the quantities 
represented by the dashed and solid lines of fig. \ref{fig:2}. 
 Since the experimental errors in 
$\tau_5^A(Q^2)$ are smaller than those associated with $R_5(Q^2)$ 
fig.~\ref{fig:2} provides a better ilustration of our results. 
Note that we obtain similar results 
for all values of $n$ ($n$=3-8) for which data is available \ci{sim1}.

The $Q^2$-dependence of n-th moment ratios and the relevance of off-shell
corrections has been recently discussed by Cothran {\it et al} \ci{sim1}.
In ref. \ci{sim1} the structure function of an on-shell nucleon, $F_2^{ON}$  
is written in terms of the relativistic invariant vertex function $\Phi(S)$  
as
\beq
F_2^{ON}(x) = \frac{x^2}{1-x} \int \frac{d^2p_t}{2(2\pi)^3}
\left[ \frac{\Phi(S)}{x^2(S-m^2)} \right]^2\ ,  
\label{def:f2sim}
\eeq
where $S = p_t^2/(x(1-x)) + \lambda^2/(1-x) + \mu^2/x$,
$\mu$ and $\lambda$ being the masses of the struck and spectator
quarks respectively \ci{sim1,close}. 
Applying Old-Fashioned-Perturbation-Theory to a system consisting 
of a constituent of mass $\mu$ and unit charge plus a neutral core
of mass $\lambda$ one finds \ci{brod}
\beq
F_2(x) = \frac{x^2}{1-x} \int \frac{d^2p_t}{2(2\pi)^3}\ 
\int d\lambda^2 \left[
\frac{ \Phi_\lambda( x {\widetilde S} ) }{ x {\widetilde S} } \right]^2\ ,
\label{def:brod}
\eeq
where ${\widetilde S} = m^2 - S$ and 
$\Phi_\lambda(x{\widetilde S})/{\widetilde S}$ can be identified with the
Fock space wave function in the infinite momentum frame. Comparison
to eq.(\ref{def:brod}) shows that eq.(\ref{def:f2sim}) 
can be obtained carrying out an integration over the squared mass 
of the spectator system, $\lambda^2$, and is strictly applicable only in the 
$x \rightarrow 1$ limit (see, e.g., ref.\ci{close}).  
Our numerical results show that the integrations involved in the 
calculations of the nuclear structure function (see eq.(\ref{def:f2aof}))
receive non negligible contributions from the region of small $(x/z)$
(for example, at $Q^2$ = 2 (GeV/c)$^2$ and $x$=0.7 more than 
50 \% of the integrated strength comes from the region $(x/z) \leq$ 0.7). 
Hence, the results of the approach of ref.\ci{sim1} should be taken with
some caution.
 
In conclusion, we have shown that the use of off-shell kinematics, appropriate 
to describe lepton scattering off bound nucleons, in the calculation of
the nuclear structure function $F_2^A(x,Q^2)$ leads to a sizable breakdown
of factorization. The amount of the effect depends upon {\em both} the form 
of nuclear spectral function {\em and} the model of the nucleon structure 
function $F_2(x,Q^2)$.  

Our approach is based on a standard nonrelativistic many-body treatment of
nuclear dynamics, involving no adjustable parameters. 
Pionic and relativistic corrections may also play a role in this context, 
but unfortunately our formalism has not yet been developed to include
their contributions in a consistent fashion.

We find that the the maximum deviation associated with off-shell effects 
is $\sim 10 \%$ and occurs at $Q^2 \sim 2$ (GeV/c)$^2$, while
at $Q^2 \geq$ 5 (GeV/c)$^2$ the effect becomes negligibly small.
Our calculations, based on a realistic spectral function and the CKMT model 
for the nucleon structure function, suggest that nuclear
effects may be larger than previoulsy estimated \ci{sim1}, and that 
 the extraction of higher twist corrections 
to $F_2(x,Q^2)$ from nuclear data may be questionable.
 
The authors gratefully acknowledge very helpful discussions with 
W. Weise.
  

\begin{figure}
\caption{
$Q^2$-dependence of $R_5(Q^2)=M^A_5(Q^2)/M^N_5(Q^2)$. The solid curve
shows the results of our approach, whereas the horizontal line 
represents the prediction of the convolution model 
(eqs.(\protect\ref{def:f2a}) and (\protect\ref{def:fact})). The data is 
taken from ref.\protect\ci{sim1}.
}
\label{fig:1}
\end{figure}


\begin{figure}
\caption{
$Q^2$-dependence of $[M^A_5(Q^2)/M^A_5(Q_0^2)]^{-1/d_5}$. The 
solid line corresponds to the full calculation, whereas the dashed 
line has been obtained neglecting off-shell corrections, i.e. 
setting $\Delta=0$ in eq.(\protect\ref{def:phi0}).    
} 
\label{fig:2}
\end{figure}

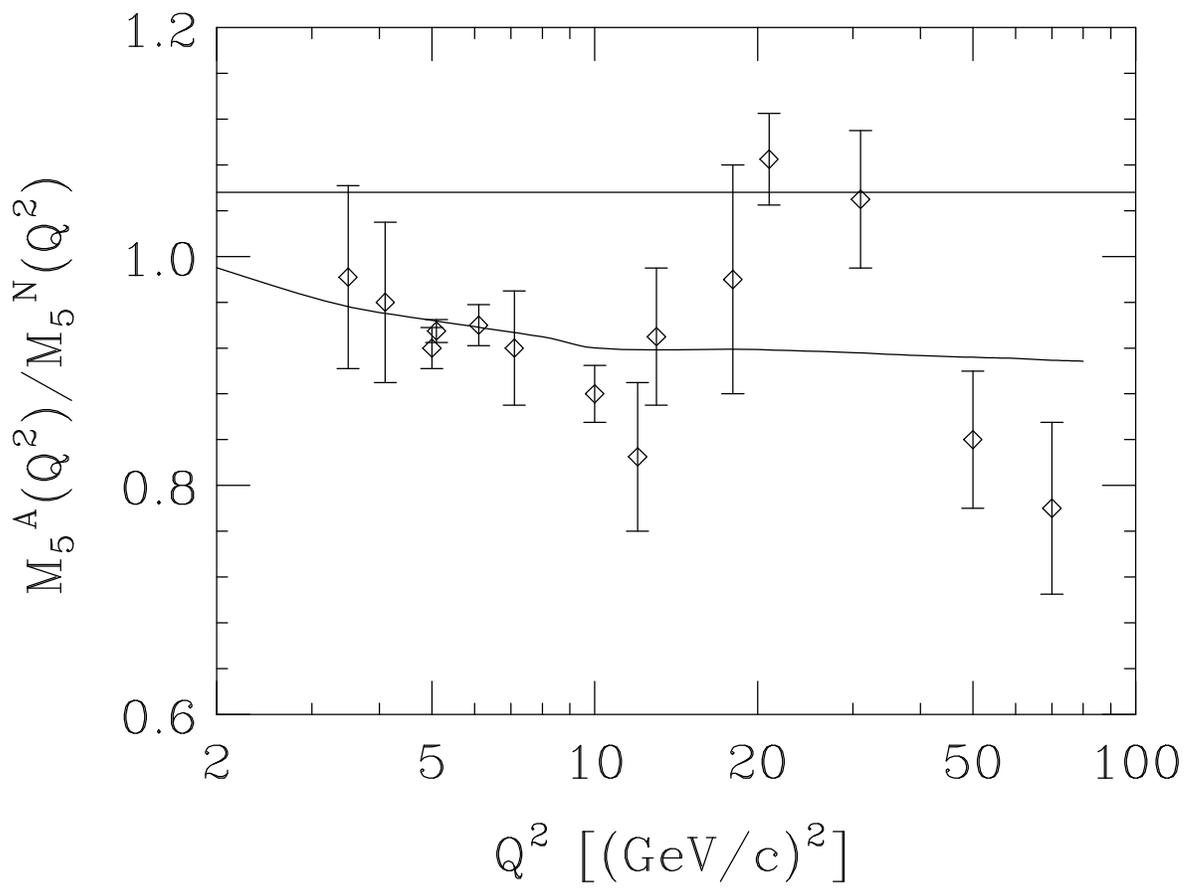
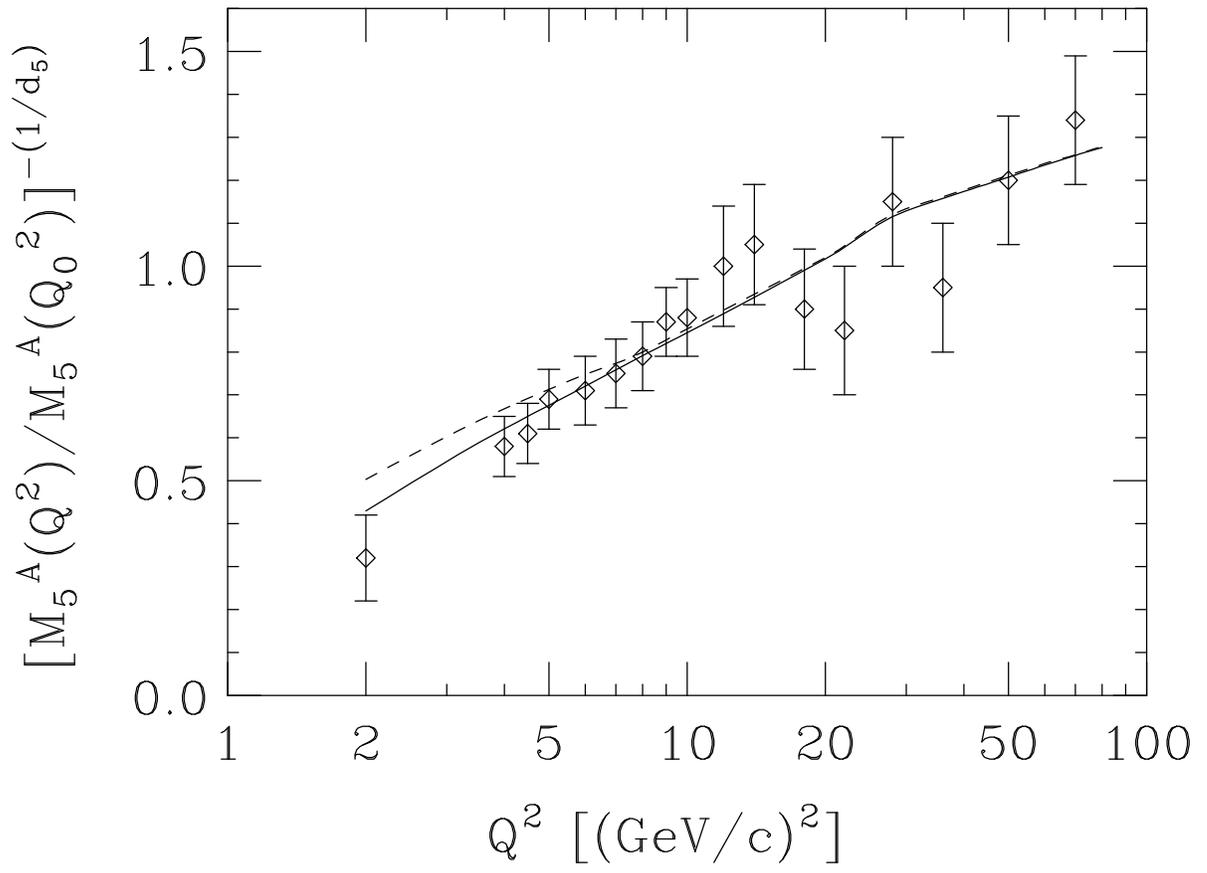
\end{document}